\documentclass[reprint,amsmath,amssymb,aps]{revtex4-2}
\newcommand{\be}{\begin{equation} } 
\newcommand{\ee}{\end{equation} } 
\newcommand{\ba}{\begin{array} } 
\newcommand{\ea}{\end{array} } 
\newcommand{\bear}{\begin{eqnarray} } 
\newcommand{\eear}{\end{eqnarray} }

\newcommand{\sev}{S_{\rm ev}}
\newcommand{\bev}{B_{\rm ev}}

\usepackage{graphicx, comment, multirow, appendix, tikz-feynman, tikz}
\usepackage{dcolumn}
\usepackage{bm}
\usepackage{hyperref}
\usepackage{lineno}
\usepackage{float}
\usepackage{cleveref}

\begin{document}

\noindent \makebox[5.cm][l]{\small \hspace*{-.2cm} }{\small IFIN-DFPE-25-0601}  
\title{
Observability of an ultraheavy diquark decaying into vectorlike quarks at the LHC
}
\author{Daniel C. Costache$^{1,2}$, Calin Alexa$^1$, Anca M. Dinu$^{1,2}$, Ioan M. Dinu$^1$, Ioana Duminica$^{1,2}$, Matei S. Filip$^{1,2}$ and Gabriel C. Majeri$^2$}
\affiliation{
$^1$Particle Physics Department, IFIN-HH, 077125 M\u agurele, RO \\
$^2$Faculty of Physics, University of Bucharest, 077125 M\u agurele, RO }
\date{\today}
\email{Corresponding author: calin.alexa@cern.ch}
\begin{abstract}
We present a comprehensive analysis of the discovery reach and exclusion limits for an ultraheavy diquark scalar (7–9.5 TeV) decaying into a pair of vectorlike quarks (1.5–2 TeV) at the HL-LHC. Building on an improved signal selection efficiency achieved using Machine Learning techniques, we extend our previous six-jet final-state study by providing a complete likelihood-based statistical treatment of this search topology. The analysis incorporates theoretical and systematic uncertainties through nuisance parameters within the likelihood framework, enabling a consistent statistical interpretation. The mass regions of interest were determined through scans of the local $p$-values, $CL_s$, and the upper limits on the model-dependent signal strength $\mu$. The results indicate a promising sensitivity to ultraheavy diquark scalars within the explored mass range, suggesting that the HL-LHC could either discover or set stringent exclusion limits on such particles.
\end{abstract}

\maketitle

\section{\label{sec:intro}Introduction}
Searches for new heavy resonances at the TeV scale remain a central objective of the ATLAS and CMS experiments at the LHC, as well as a major focus of collider phenomenology \cite{CMS:2022usq, ATLAS:2023ssk, CMS:2025hpa, Crivellin:2022nms}. The upcoming ATLAS \cite{CERN-LHCC-2015-020} and CMS \cite{Butler:2055167} detector upgrades further motivate searches for new physics at the multi-TeV scale \cite{Dobrescu:2019nys}, where both experiments have already explored ultraheavy resonances \cite{ATLAS:2023ssk,ATLAS:2024gyc,CMS:2022usq,CMS:2022fck,CMS:2024ldy,CMS:2025hpa}. 

A recent CMS analysis \cite{CMS:2025hpa} of broad ultraheavy resonances in four-jet final states reinterprets an earlier narrow-resonance search \cite{CMS:2022usq} by extending the resonance width to 10\%. The reported global significance for a four-jet resonance at 8.6~TeV ranges between 3.9$\sigma$ and 3.6$\sigma$. Both studies interpret their results within a Beyond Standard Model (BSM) framework featuring a diquark scalar coupling to up quarks and same-sign vectorlike quarks (VLQ) 
\cite{Dobrescu:2018psr,Dobrescu:2024mdl}.

In our previous study \cite{Duminica_2025}, we analyzed the discovery potential of an ultraheavy diquark scalar, $S_{uu}$, at the HL-LHC in the $7-8.5$ TeV mass range. Owing to its couplings to two up quarks, such a resonance can be produced with sizable cross section even near the kinematic limit of the collider. We consider the six-jet final state from $pp\to S_{uu}\to\chi\chi\to (W^+b)(W^+b)\rightarrow (jjb)(jjb)$, where $\chi$ is a vectorlike quark. We show that, with an integrated luminosity of $3000\ \text{fb}^{-1}$, searches at the LHC could discover or exclude $S_{uu}$ masses up to about 8 TeV for couplings as small as $y_{uu} \approx 0.2$.

Extending our previous phenomenological study, this work provides the first comprehensive likelihood-based statistical analysis of the six-jet final state in this model, enabling a quantitative evaluation of its discovery and exclusion prospects at the High-Luminosity LHC (HL-LHC) \cite{ZurbanoFernandez:2020cco}.

Complementing our earlier results, which focused on event generation, kinematic characterization and Machine Learning-based signal selection, we develop a full statistical framework incorporating nuisance parameters, pseudo-experiments, and the computation of local $p$-values, $CL_s$ and 95\% confidence level limits. 

The analysis also includes a careful treatment of theoretical and systematic uncertainties within the likelihood framework. 

While the underlying physics scenario and qualitative phenomenology remain unchanged, this approach provides a quantitatively controlled determination of the discovery reach and exclusion potential.    

Complementary channels are being investigated in parallel, including single production \cite{Filip:2601.11181} $S_{uu}\to u\chi$ and pair-production mode $S_{uu}\to\chi\chi$ with $\chi\to tZ, ~th^0$ in fully hadronic final states. A combined interpretation will be addressed in future work.

The paper is organized as follows. \Cref{sec:theory_data} reviews the BSM framework, where the relevant Lagrangian terms and the main theoretical assumptions are explicitly summarized. It describes the Monte Carlo (MC) generation of the data samples and the Machine Learning (ML) method employed for signal–background discrimination and discusses the effect of a 1.5~TeV vectorlike quark mass hypothesis on signal selection efficiency. The treatment of theoretical and systematic uncertainties is discussed in \Cref{sec:uncert}, where their propagation to the statistical analysis is described in \Cref{sec:results}. The statistical framework used to evaluate the discovery potential and exclusion limits is presented in \Cref{sec:stats}, while the corresponding results, including the impact of these uncertainties on the local $p$-values and the derived exclusion limits, are discussed in \Cref{sec:results}. Finally, \Cref{sec:conclusions} summarizes the main findings and outlines future prospects.

\section{\label{sec:theory_data}BSM framework and data processing methods}

Vectorlike quarks ($\chi$) are color-triplet fermions with electric charge $+2/3$, while $S_{uu}$ is a color-sextet complex scalar with electric charge $+4/3$. The introduction of these two BSM particles, $S_{uu}$ and $\chi$, into the Standard Model (SM) Lagrangian is thoroughly discussed in \cite{Dobrescu:2018psr,Dobrescu:2019nys,Dobrescu:2024mdl}. Thus, the relevant extra Lagrangian terms explicitly show the interactions between the diquark scalar, the vectorlike quarks and the SM up quarks:
\begin{equation}\label{eq:lagrangian}
    \frac{y_{\chi\chi}}{2} S_{uu}  \overline \chi_{R} \chi^c_{R} +  y_{u\chi} S_{uu}  \overline u_{R} \chi^c_{R} + \frac{y_{uu}}{2} S_{uu}  \overline u_{R} u^c_{R} + {\rm H.c.},
\end{equation}
with $y_{uu}$, $y_{u \chi}$ and $y_{\chi \chi}$ being the corresponding Yukawa couplings and the $c$ index accounting for the charge conjugation. The $1/2$ factor is present for the $\chi\chi$ and the $uu$ terms as to avoid double counting.

In the invariant-mass window centered at $M_S=8$ TeV with a $4\%$ width \cite{Dobrescu:2018psr}, the event yield is highly dominated by the $s$-channel contribution. The $t$-channel VLQ pair-production contribution is negligible in this region. For this reason, the present analysis focuses on the resonant $s$-channel topology, where the on-shell peak provides the dominant signal fraction.

This theoretical framework proposes that the diquark scalar has three decay modes, with the branching ratios depending on Yukawa couplings, vectorlike quark and diquark scalar masses. For $m_\chi = 2$ TeV and $M_S=8$ TeV, in the conservative scenario for the Yukawa couplings ($y_{uu}=0.2$, $y_{u \chi}=0.1$ and $y_{\chi \chi}=0.3$), the corresponding branching ratio for each decay mode is: $B(S_{uu}\rightarrow \chi \chi)=56.5\%$, $B(S_{uu}\rightarrow u \chi)=13.7\%$ and $B(S_{uu}\rightarrow uu) = 29.8\%$. Vectorlike quarks are considered to decay into SM boson-fermion pairs, resulting in three distinct channels, with the branching ratios: $B(\chi\rightarrow Wb)=50\%$, $B(\chi\rightarrow Zt)=B(\chi\rightarrow h^0t) = 25\%$.

We choose model parameters such that the renormalizability of the theory is preserved. The diquark scalar mass range is chosen between 7 and 9.5 TeV, with 0.25 TeV steps. Current bounds on diquark scalars are set at 7.3 TeV, for $95\%$ C.L. in dijet searches \cite{CMS:2018wxx}. All three Yukawa couplings have subunitary values and the $y_{\chi \chi}/y_{uu}$ fraction is set at 1.5. Two parameters benchmark scenarios are considered, corresponding to the product of the $S_{uu}$ Yukawa coupling to up quarks and the branching fraction of the VLQ decay to $W^+b$:
\be
y_{uu} B(\chi \to W b) = 0.1 \, ,\; 0.2 ~. \label{eq:benchmark_sensitivity}
\ee
These correspond to coupling values of $y_{uu} = 0.2$ and $0.4$ respectively, while $y_{\chi\chi}=1.5y_{uu}$ \cite{Dobrescu:2018psr}.

Data samples for the six-jet final state of the signal processes are generated with \textsc{MadGraph5\_aMC@NLO} (v3.3.2) \cite{Alwall:2014hca} following \cite{Duminica:2024nos}. Showering and hadronization are performed with \textsc{Pythia8.310} \cite{Bierlich:2022pfr} using the NNPDF23LO \cite{NNPDF:2014otw} set. Jets are reconstructed with the anti-$k_{t}$ algorithm \cite{Cacciari:2008gp} in \textsc{FastJet} \cite{Cacciari:2011ma}. Multi-parton interactions and initial- and final-state radiation are included. Detector effects are simulated with the ATLAS configuration in \textsc{Delphes 3.5.0} \cite{deFavereau:2013fsa}. Background processes are generated with \textsc{Pythia8.310} in high phase spaces using the minimum invariant mass variable $mHatMin$ \cite{Bierlich:2022pfr} (which we will denote with $\widehat{m}_{\text{min}}$ in the following paragraphs). In total, 32 background event types are included in our analysis, from the following compressed categories: $2\rightarrow 2$ QCD, $W+$jets, Higgs processes, dibosons, and $t\bar t$. The complete list is to be found in \cite{Duminica:2024nos, Git:2023V2}.

Compared to our previous study \cite{Duminica_2025}, a significantly larger set of Monte Carlo (MC) samples was generated at $\sqrt{s}=13.6$ TeV for both signal and background processes. Signal events were simulated in the mass range $M_S\in [7,9.5]$ TeV in steps of 0.25~TeV, while the corresponding background samples were produced with a phase-space cut of $\widehat{m}_{\text{min}} = M_S - 0.5~\text{TeV}$ for each $M_S$ value.

Signal-from-background selection is performed using the Random Forest Machine Learning algorithm in the \textsc{SCIKIT-LEARN} implementation~\cite{Pedregosa:2011}. The source code is accessible at \cite{Git:2023}. The ML input consists of 75 jet-based kinematic and topological observables, including single-jet kinematics, all dijet and trijet invariant masses and angular separations, as well as global event quantities such as jet and b-tag multiplicities and reconstructed W-candidate counts. Based on these variables, the model is trained on $80\%$ of the data. The remaining $20\%$ of the data is used for testing, where the RF outputs, for each event, a classifier score interpreted as the probability of being signal. This score defines the discriminator D, which is used to select events by applying a threshold on its value. The final event yields are obtained by weighting the simulated samples with their corresponding cross section, integrated luminosity and selection efficiencies, denoted as $\sev$ (signal) and $\bev$ (background). 

The performance and associated uncertainties of the algorithm are firstly evaluated using the k-fold cross-validation technique. Secondly, the classifier performance is assessed by varying selected hyperparameters and studying their impact on signal efficiency and background rejection. The parameters considered include the number of decision trees, the maximum tree depth, the minimum number of samples required for internal node splitting and the minimum number of samples per leaf \cite{10.1007/978-3-642-31537-4_13, Hastie:2009, Probst:2019}. The random seed is fixed, to assure the analysis reproducibility.

To assess the stability of the RF performance, three representative configurations are considered by varying key hyperparameters, as listed in \Cref{tab:rf_hyperparameters}. The default settings (Model 0) have been used in \cite{Duminica_2025}. 
\begin{ruledtabular}
\begin{table}[h]
    \centering
    \caption{Three RF hyperparameter configurations.}
    \label{tab:rf_hyperparameters}
    \begin{tabular}{l|ccc}
RF hyperparameter & Model 0 & Model 1 & Model 2 \\[1mm] \hline \\[-2.6mm]
No of estimators & 100 & 250 & 400 \\[1mm]
Maximum tree depth & 0 & 10 & 12 \\[1mm]
Minimum samples for split & 2 & 50 & 20 \\[1mm]
Minimum samples per leaf & 1 & 20 & 5  \\[1mm]
Random seed & 42 & 42 & 42 \\[1mm]
    \end{tabular}
\end{table}
\end{ruledtabular}

The number of trees is chosen in the range 100–400 to ensure a sufficient ensemble averaging while controlling computational cost. The maximum tree depth is varied between unrestricted growth and moderate depths (10–12), allowing us to probe the balance between model complexity and overfitting. The minimum number of samples required for node splitting and for leaf nodes is also varied, from minimal values that favor detailed partitioning of the feature space to larger values that enforce smoother decision boundaries. These variations enable a controlled study of how model complexity affects signal efficiency, background rejection, and overall classification performance. The resulting variations are used to evaluate the robustness of the classifier response and its impact on the statistical sensitivity of the analysis.

To select the optimal classifier configuration for each sample, we adopt an evaluation strategy based on physics-driven observables rather than standard ML metrics. Instead of relying on accuracy or AUC, which do not account for event weights, the selection is performed using the weighted signal and background yields evaluated at three working points, $D=0.80,~0.90$ and 0.925. These thresholds probe the high-score region, where QCD background contamination is most relevant.  

For each model and threshold, we define the combined score
\be
Z_{\text{score}}(D) = \frac{S_{\text{ev}}(D)}{\sqrt{B_{\text{ev}}(D)}} \cdot \frac{1}{1 + \sigma_{B_{\text{ev}}(D)}/B_{\text{ev}}(D)}
\ee

which favors configurations with high signal significance while penalizing large fluctuations in the background estimate across k-folds. To ensure robustness, configurations with $\sigma_{B_{\text{ev}}(D)}/B_{\text{ev}}(D) > 0.3$ at any of the considered thresholds are discarded. This requirement removes models exhibiting unstable background behavior in the high-discriminator region. For each retained configuration, the final performance metric is defined as the average score over the three thresholds. The optimal configuration for each sample was then selected by maximizing this average score. This ensures both strong signal-to-background discrimination and stable behavior across the relevant thresholds. 

The CMS~\cite{CMS:2022fck} lower mass limit for pair-produced VLQs is approximately 1.4~TeV for the branching ratio scenario $B(\chi \rightarrow W b$):$B(\chi \rightarrow h^0 t$):$B(\chi \rightarrow Z t$)=$50\%$:$25\%$:$25\%$. A more recent study published by the ATLAS Collaboration excludes the production of vectorlike quarks with masses lower than 1.36 TeV ~\cite{ATLAS:2024gyc}. Motivated by these results, we study the impact of lowering $m_{\chi}$ to 1.5 TeV.

\begin{figure}[H]
    \centering
    \includegraphics[width=0.5\textwidth]{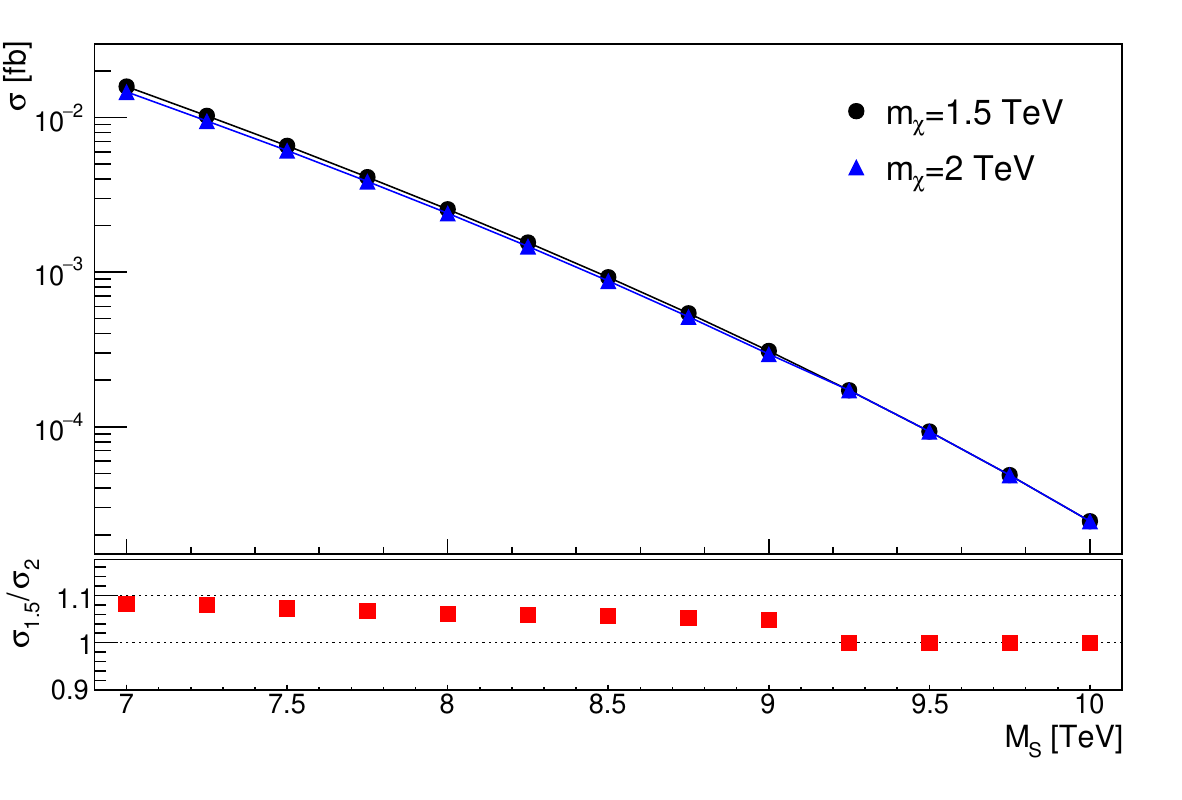}
    \caption{\label{fig:cross-section}Cross-section for $pp\to S_{uu}\rightarrow \chi\chi$ at $\sqrt{s}=13.6$ TeV, when $\chi \rightarrow W^+b \rightarrow (jj)b$, $y_{uu}=0.2$ and $m_\chi = 1.5$ and 2 TeV.}
\end{figure}

We observe an approximately $10\%$ increase in cross section for $m_{\chi}=1.5~\text{TeV}$ relative to $m_\chi=2~\text{TeV}$, persisting up to $M_S \leq 9$ TeV, as shown in \Cref{fig:cross-section}. Within the narrow-width approximation~\cite{Dobrescu:2018psr}, the six-jet final state cross section is strongly dependent on $M_S$, reflecting the dominance of the on-shell resonance contribution.

The VLQ mass impacts the final-state kinematics, including jet transverse momenta and event topology, thereby affecting the signal selection efficiency, background discrimination and statistical sensitivity. The dependence of the signal ($S_{\text{ev}}$) and background ($B_{\text{ev}}$) event yields on the multivariate discriminator threshold (D), obtained for $M_S=8~\text{TeV}$, $m_\chi = 1.5$ and $2~\text{TeV}$, $\widehat{m}_{\text{min}}=M_S - 0.5~\text{TeV}$, 
$\sqrt{s} = 13.6~\text{TeV}$, 
and $y_{uu}B(\chi\to Wb) = 0.1$, is summarized in \Cref{tab:discriminator_values}.
\begin{ruledtabular}
\begin{table}[h]
\caption{\label{tab:discriminator_values}
$\sev$ and $\bev$ yields dependence on $D$ for $M_S=8$ TeV, $m_{\chi}=1.5, \ 2$ TeV, $\widehat{m}_{\text{min}}=M_S - 0.5$ TeV, $\sqrt{s}=13.6$ TeV, $y_{uu}B(\chi \rightarrow Wb)=0.1$. $\sev$ errors are below 1\%.
}
\centering
\footnotesize
\begin{tabular}{l|ccccc}
& $D=0.80$ & $D=0.90$ & $D=0.925$ & $D=0.95$ & $D=0.96$ \\[1mm]   
\cline{2-6}
\\[-1.6mm]
\multicolumn{6}{c}{$m_\chi=1.5$ TeV} \\[1mm]
\hline \\[-2.6mm]
$\sev$ & 7.64 & 7.62 & 7.60 & 7.48 & 7.25 \\[1mm]
$\bev$ & 8.49{\tiny$\pm$}0.49 & 3.09{\tiny$\pm$}0.36 & 1.90{\tiny$\pm$}0.26 & 0.73{\tiny$\pm$}0.24 & 0.41{\tiny$\pm$}0.23 \\[1mm]
\hline \\[-1.6mm]
\multicolumn{6}{c}{$m_\chi=2$ TeV} \\[1mm]
\hline \\[-2.6mm]
$\sev$ & 7.20 & 7.18 & 7.10 & 6.95 & 6.35 \\[1mm]
$\bev$ & 5.34{\tiny$\pm$}0.44 & 1.68{\tiny$\pm$}0.56 & 0.67{\tiny$\pm$}0.12 & 0.36{\tiny$\pm$}0.12 & 0.09{\tiny$\pm$}0.07 \\[1mm]
\end{tabular}
\end{table}
\end{ruledtabular}

For $m_\chi = 2~\text{TeV}$, the background yield $B_{\text{ev}}$ falls below one at lower discriminator thresholds than in the $m_\chi = 1.5~\text{TeV}$ scenario, while the signal yield remains approximately constant between the two cases. This behavior reflects the harder kinematic features of the heavier vectorlike quark, which lead to more distinctive signal topologies and improved background suppression as the discriminator threshold increases.

\section{Uncertainties} \label{sec:uncert}

Background uncertainties are evaluated only for the dominant components, which together account for more than $90\%$ of the total background yields. The leading contributions arise from $h^0+jj$, $W+j$, and $q\bar{q}\to gg$, whose combined fraction increases from $\approx90\%$ at $M_S = 7~\text{TeV}$ to $\approx98\%$ at $M_S = 8.75~\text{TeV}$.

Theoretical uncertainties for both signal and background, arising from renormalization- and factorization-scale variations as well as PDF uncertainties \cite{10.21468/SciPostPhysCore.6.2.045}, are estimated with \textsc{MadGraph5\_aMC@NLO} (v3.7.0) \cite{Alwall:2014hca}. 

A dynamical scale choice based on the partonic center-of-mass energy is used, and both the renormalization and factorization scales are varied by a factor of two following the standard seven-point prescription, in which the samples are evaluated for the seven $(\mu_R, \mu_F)$ combinations: $(0.5,1)$, $(1,0.5)$, $(1,2)$, $(2,1)$, $(0.5,0.5)$, $(1,1)$ and $(2,2)$, where $(1,1)$ defines the nominal choice. We find that the scale uncertainty is asymmetric for both signal and background samples. For the signal process over the mass range $M_S\in [7,\ 9.5]$, the variations span from $^{+19.1\%}_{-15.4\%}$ to $^{+26.3\%}_{-20.6\%}$. For the background processes, the size of the uncertainty depends strongly on the event type. The smallest variations are obtained for the $W+$jets processes, ranging from $^{+31.7\%}_{-20.7\%}$ to $^{+33.6\%}_{-21.0\%}$, depending on the chosen $\widehat{m}_{\text{min}}$. By contrast, the largest scale uncertainties are found for the Higgs backgrounds, varying between $^{+36.7\%}_{-21.7\%}$ and $^{+43.7\%}_{-22.5\%}$.

For signal samples with $M_S\in [7,\ 9.5]$ TeV, the PDF uncertainty ranges from $\pm 9.64\%$ to $\pm 21.5\%$.  The background sources are even more sensitive to PDF variations, mainly because of the high-invariant-mass phase space cut applied at generation level (see \Cref{sec:theory_data}). The corresponding theoretical uncertainties were evaluated separately for individual background process with \textsc{MadGraph5\_aMC@NLO} (v3.7.0) \cite{Alwall:2014hca}. Depending on the process and $\widehat{m}_{\text{min}}$, PDF uncertainties in the background yields range from about $\pm27.8\%$ for $W+$jets events with $\widehat{m}_{\text{min}}=7.5$ TeV to values as large as $\pm85.1\%$ for $q\bar{q}\to gg$ at $\widehat{m}_{\text{min}}=9$ TeV.

Systematic uncertainties were evaluated for the signal and background simulations at $M_S=8~\text{TeV}$.

Pile-up effects were evaluated for scenarios with 25, 50, 100, 150, and 200 interactions per bunch crossing, leading to uncertainties of up to $3.5\%$.

To evaluate the impact of jet energy scale (JES) and jet energy resolution (JER) uncertainties, the signal and background samples were re-simulated under several configurations. JES and JER were varied independently by scaling each quantity up and down by factors of two and one-half, and additional correlated variations were considered by simultaneously increasing or decreasing both quantities by the same factors. These variations define a multi-point scheme analogous to the standard seven-point prescription used for theoretical uncertainties. The sensitivity to jet reconstruction was further examined by varying the jet clustering radius parameter in the anti-$k_t$ algorithm. While R=0.6 was taken as the baseline choice, additional configurations with R=0.2 and R=1.0 were used to probe the dependence of the event selection on the jet topology. Overall, the detector-related uncertainties are dominated by jet reconstruction effects ($6.9\%$) and by JES/JER variations ($6.8\%$).

The HL-LHC luminosity is estimated to be measured with $1\%$ precision \cite{Giacobbe_2024}.

Uncertainties from the ML-based event selection, evaluated using k-fold cross-validation, are propagated to the statistical analysis.

\section{\label{sec:stats}Statistical methods}

The theoretical and systematic uncertainties described above are propagated to the statistical interpretation in order to evaluate their impact on the discovery and exclusion sensitivities. Variations associated with renormalization- and factorization-scale choices, PDF uncertainties, and detector-related effects are incorporated in the expected signal and background yields. The resulting shifts in the local $p$-value and in the derived $CL_s$ exclusion limits are then examined to assess the robustness of the analysis in the presence of these uncertainties.

Although some of the individual theoretical uncertainties become large in the high-mass regime, this does not invalidate the statistical interpretation. Rather, it indicates that the resulting sensitivities should be viewed as baseline estimates subject to sizable normalization effects. The purpose of the nuisance treatment adopted here is therefore to quantify the stability of the $p$-values and exclusion limits under realistic variations of the dominant background components, rather than to provide an experiment-level precision uncertainty model.

The statistical interpretation is performed using a likelihood function in which nuisance parameters are assigned only to leading background components, while subleading backgrounds are neglected due to their insignificant contribution after the RF selection. This ensures a consistent treatment of the relevant uncertainties while maintaining a minimal and robust statistical model.

To draw robust conclusions from the results obtained with the RF-based signal selection, an extensive statistical analysis was carried out using methods consistent with those adopted by the ATLAS and CMS Collaborations~\cite{CMS-NOTE-2011-005,Read_2002}. Two independent analysis paths were pursued, each relying on a distinct computational framework. The \textsc{RooFit}~\cite{roofit} toolkit, implemented within the \textsc{ROOT} framework (v6.32.10)~\cite{root}, was employed to compute local $p$-values and $CL_s$ metrics, with parallelization achieved via the \textsc{TProcessExecutor} class. The \textsc{RooStats}~\cite{roostatsproject} package, which provides a higher-level interface to \textsc{RooFit}, was used to perform the upper-limit scans~\cite{Git:statistics}.

The workflow followed in the \textsc{RooFit} program consists of taking $S_{\text{ev}}$ and the event counts for the dominating background processes mentioned in \Cref{sec:uncert}, along with their uncertainties and building a probability density function (\textit{pdf}, to be differentiated from the parton distribution function PDF). The Poisson term for the number of events is defined as:
\begin{equation}
    \label{eq:likelihood-poisson}
    \mathcal{L}_{\text{Pois}}\left(N_{\text{obs}}\mid N_{\text{exp}}\right) = \text{Pois}(N_{\text{obs}}\mid N_{\text{exp}}),
\end{equation}
where $N_{\text{obs}}$ is the observed number of events and the expected number of events is defined as: 
\begin{equation}
   N_{\text{exp}} =\mu\times S_{\text{ev}} \times \prod_i \theta_{S,i} +\sum_{k}B_{\text{ev},k} \times \prod_j \theta_{B_k,j}.
\end{equation}
Here, $\mu$ is the signal strength multiplier, $\theta_S$ are the uncertainty sources for the signal and $\theta_{B_k}$ are the uncertainty sources for each background process. These account for luminosity uncertainties, pile-up, jet energy resolution and jet energy scale effects, renormalization/factorization scale and PDF variations, as well as uncertainties coming from the k-fold algorithm. All of them are presented quantitatively in \Cref{sec:uncert}. A pseudo-observed event count, $N_{\text{obs}}$, is generated under the specified hypothesis and used throughout this section.

To constrain the fitting of the nuisance parameters in the model, log-normal terms are applied instead of Gaussian ones, in order to avoid issues that can arise with positively defined observables, as discussed in Ref.~\cite{CMS-NOTE-2011-005}. These constraints are defined around the nominal value of each parameter, with the scale factor of the log-normal being built as $\kappa = 1+\sigma_{\text{rel}}$, where $\sigma_{\text{rel}}$ is the relative error for each aforementioned quantity. The only exception is the case of the theoretical scale variations, which present asymmetrical uncertainties, being treated by employing a split Gaussian. The total probability density function (likelihood) is then given by the product of the Poisson term and the corresponding constraint terms:
\begin{equation}
    \mathcal{L}=\mathcal{L}_{\text{Pois}}\times \prod_i \text{Constr}(\theta_i).
\end{equation}

The probability density function described above is used to generate a large number of pseudo-experiments under the background-only ($b$-only) hypothesis, where the signal strength parameter $\mu$ is set to zero. These pseudo-experiments serve as ``observed data'' for subsequent fits. The same \textit{pdf} is then fitted to each pseudo-experiment for $\mu = 1$ and for $\hat{\mu}$, the value of the signal strength estimator that maximizes the likelihood. The test statistic $\tilde{q}_{\mu}$ is defined as the negative logarithm of the likelihood ratio, as described in Ref.~\cite{Cowan_2011}. For each pseudo-experiment a large number of toy datasets are generated, both under $s+b$ and $b$-only hypotheses, in order to compute
\begin{equation}
    \label{eq:clsb_clb}
    \begin{split}
        CL_{s+b}=&P\left(\tilde{q}_{\mu}^{toy}\geq \tilde{q}_{\mu}^{\text{obs}}\mid s+b\right) \\
        CL_{b}=&1-P\left(\tilde{q}_{\mu}^{toy}\geq \tilde{q}_{\mu}^{\text{obs}}\mid b\right).
    \end{split}  
\end{equation}
These probabilities are obtained by integrating the test statistic distribution tail, starting from the value of $\tilde{q}_{\mu}^{\text{obs}}$. At the end of each pseudo-experiment, we compute $CL_{s}$ \cite{Read_2002}. A distribution of $CL_{s}$ values is constructed and the median expected values, along with the standard deviations are extracted.

Using this same pseudo-experiment method, we determined the local $p$-value in the discovery hypothesis by means of the $q_0$ test statistic, also defined in Ref.~\cite{Cowan_2011}. In comparison to the exclusion case, the pseudo-experiments are generated under $s+b$ hypothesis and the toys under $b$-only. Computing $p$-values in this way is more rigorous and reliable, as it properly accounts for the nuisance parameters in the model, in contrast to simple Poisson-based estimations. A known limitation of this approach is that when the numerical $p$-value falls below the inverse of the total number of pseudo-experiments, it may appear as zero (for example, if $10^7$ toys are generated and $p \leq 10^{-7}$, the result may be reported as $p = 0$). In such cases, where this limitation occurs, the local $p$-value was instead computed analytically using Poisson statistics. The affected points are explicitly indicated in the results section.

The hypothesis inversion used to derive the 95\% upper limits was performed in \textsc{RooStats} using a simplified implementation. The likelihood function is identical to that described previously. The \textsc{AsymptoticCalculator} class in \textsc{RooStats} is used to scan over the signal strength parameter $\mu$ and to determine the value at which the 95\% confidence level is reached, following the asymptotic formulae of Ref.~\cite{Cowan_2011}. The median expected and observed upper limits, together with the corresponding $\pm1\sigma$ and $\pm2\sigma$ uncertainty bands, are then extracted from the resulting distributions.

\section{\label{sec:results}Observability limits}

The results of the statistical analysis are encouraging across a wide $M_{S}$ range and for both $m_{\chi}$ and Yukawa coupling considered. The metrics used to assess the model performance and the quality of the results include the local $p$-value for discovery, the $CL_s$ criterion, and the 95\% upper limit on the signal strength parameter. 

Since, in some mass regions and for some discriminator values, the background yields fall below unitary, we set the discriminator threshold as stringent as possible to maximize the signal purity while maintaining event counts at or above unity. This corresponds to $D=0.90$. The preferred approach when dealing with very low event counts is to use toy-based computations, which are heavy on processing resources. 

The discovery $p$-value quantifies the probability that the background alone could produce an excess as large as the observed signal. 

The $CL_s$, defined as the ratio of the quantities in \Cref{eq:clsb_clb}, measures the probability for the signal hypothesis to exhibit background-like fluctuations and is used to determine exclusion regions in the model parameter space.

\subsection{\label{subsec:pvalue}Local $p-$values}

After computing the local $p$-values at each mass point for different discriminator thresholds, VLQ masses, and $y_{uu}$ coupling values, we observed a clear trend in the evolution of the corresponding probabilities. 

\begin{figure}[h]
\centering
\includegraphics[width=0.5\textwidth]{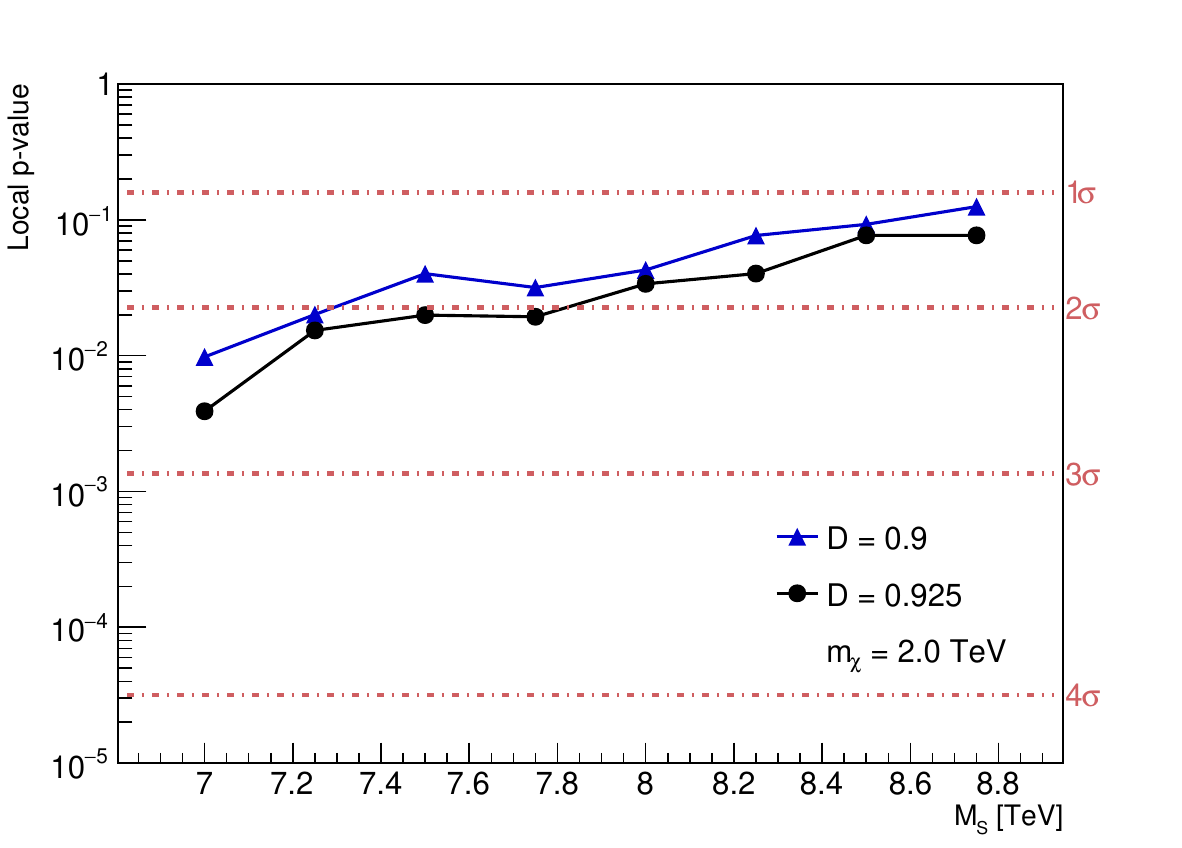} 
\caption{Local $p$-values quantifying the excess above the observed values of the test statistic $q_{0}^{\text{obs}}$, for two D values.}
\label{fig:pval_D}
\end{figure}

\Cref{fig:pval_D} shows the dependence of the estimated $p-$values on the diquark scalar mass for two discriminator thresholds, assuming $m_\chi = 2~\text{TeV}$. At $M_S=7~\text{TeV}$, the $p-$values fall below the $3\sigma$ threshold for $D=0.925$, providing evidence against the background-only hypothesis. Increasing the discriminator threshold improves the signal–background separation, leading to smaller $p-$values. We therefore adopt $D=0.90$ for the statistical analysis, as it provides a conservative and statistically stable choice.

\begin{figure}[h]
\centering
\includegraphics[width=0.5\textwidth]{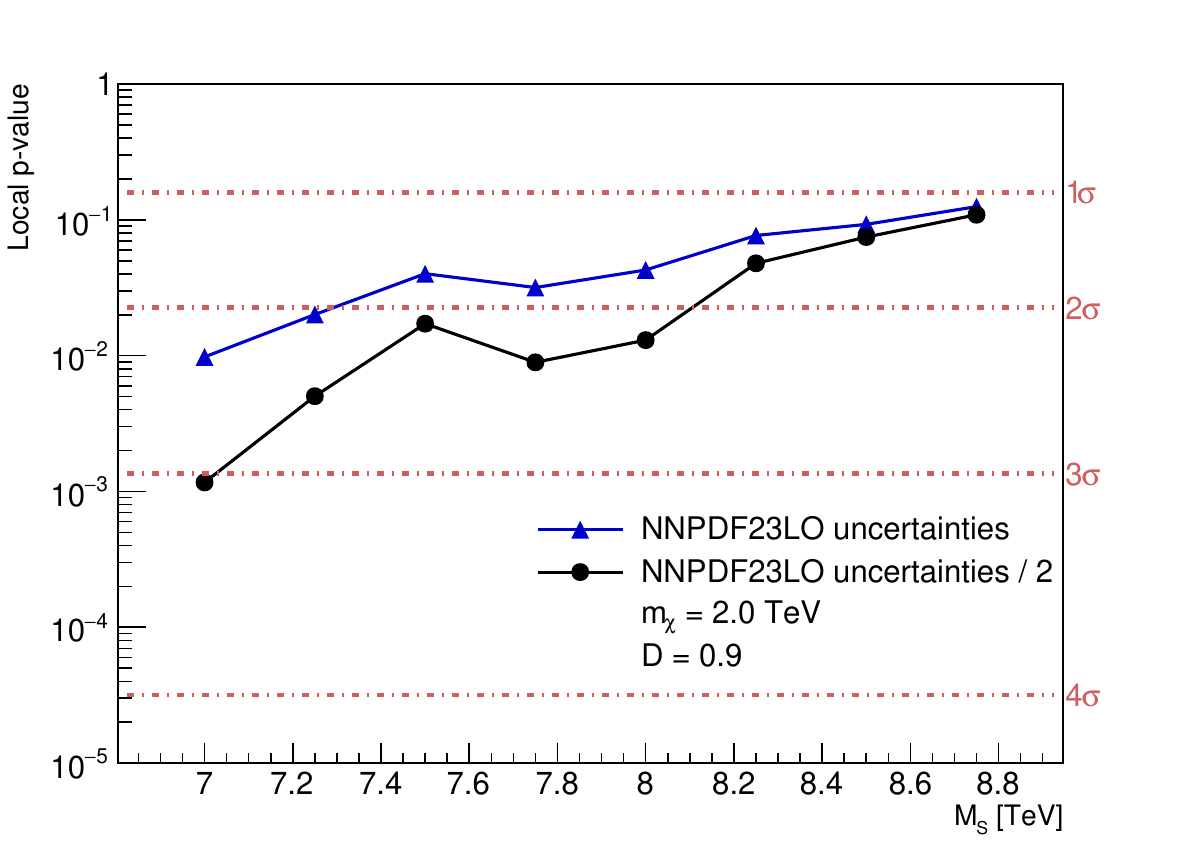} 
\caption{Local $p-$value for the observed test statistic $q_{0}^{\text{obs}}$ for two PDF uncertainty scenarios.}
\label{fig:pval_PDF}
\end{figure}

At large parton momentum fractions, corresponding to the multi-TeV regime, parton distribution functions (PDFs) exhibit sizable uncertainties, limiting predictive accuracy near the kinematic boundary of the collider~\cite{Accardi:2023pdf,Butterworth:2015oua}. To assess the impact of future improvements at the HL-LHC, we consider an illustrative scenario in which PDF uncertainties are reduced by a factor of two, consistent with expected gains in global PDF determinations~\cite{Khalek2018TowardsUP}. As shown in \Cref{fig:pval_PDF}, this leads to a modest improvement in sensitivity at lower $M_S$, with local significances approaching the $3\sigma$ level. At higher masses, the curves converge, as the PDF uncertainties, although reduced, remain sizable.

\begin{figure}[h]
\centering
\includegraphics[width=0.5\textwidth]{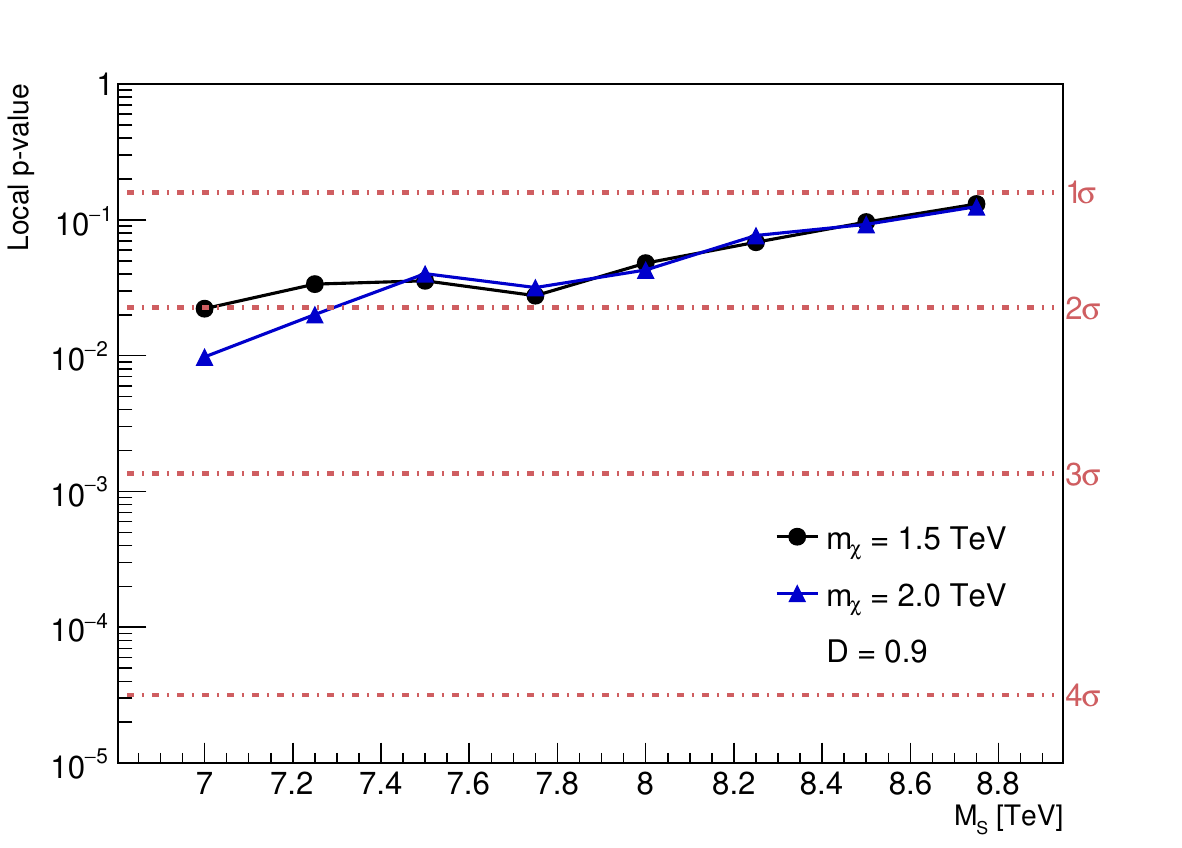} 
\caption{Local $p$-value quantifying the excess above the observed values of the test statistic $q_{0}^{\text{obs}}$, for two $m_\chi$ values.}
\label{fig:pval_mChi}
\end{figure}

With the discriminator fixed at $D = 0.9$, we evaluate the local $p-$values for $m_\chi=1.5$ and $2~\text{TeV}$, as shown in ~\Cref{fig:pval_mChi}. 

\begin{figure}[h]
\centering
\includegraphics[width=0.5\textwidth]{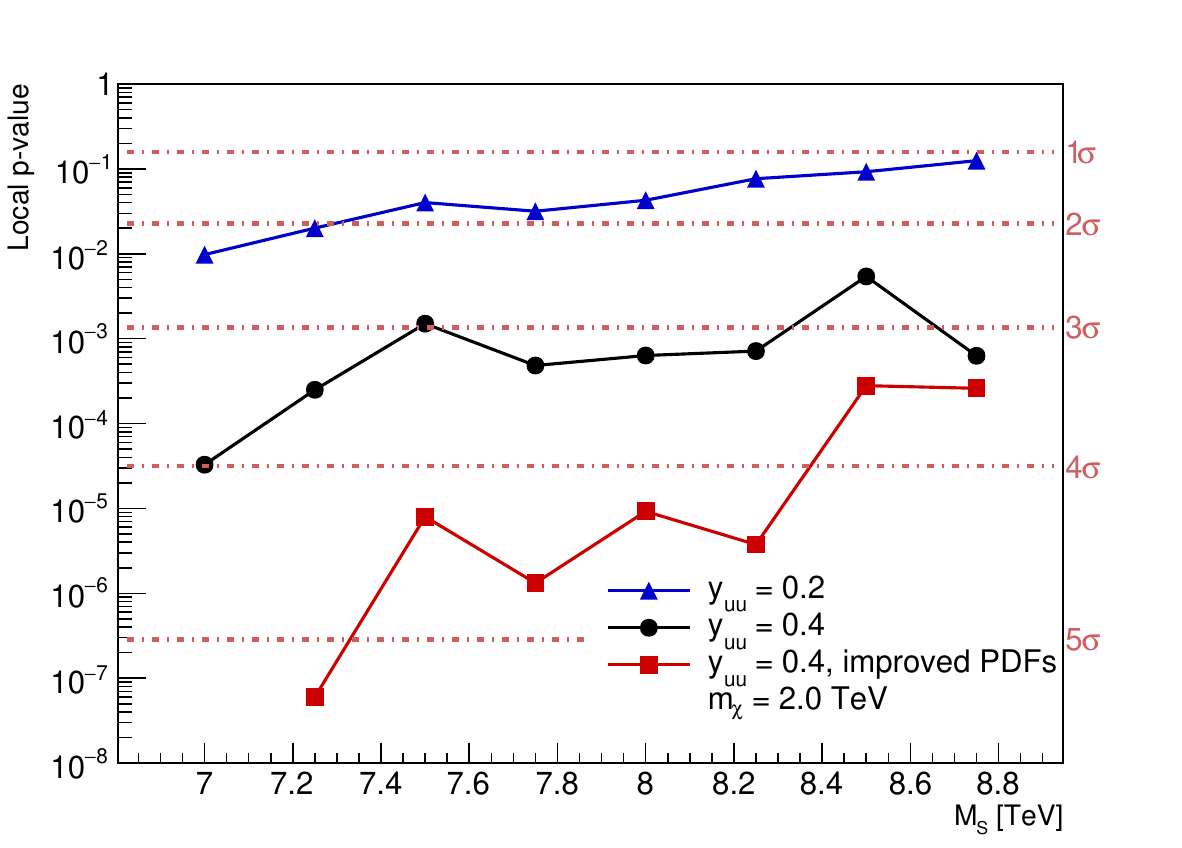} 
\caption{Local $p$-value for the observed test statistic $q_{0}^{\text{obs}}$ for two $y_{uu}$ values and a scenario with increased coupling and halved PDF uncertainties.}
\label{fig:pval_yuu}
\end{figure}

The lower VLQ mass shows reduced sensitivity at lower $S_{uu}$ masses, but at higher $M_S$ values it has virtually identical performances to the $m_\chi = 2~\text{TeV}$. This behavior can be attributed to the interplay between phase-space effects and the resulting changes in event kinematics, which affect the signal selection efficiency and background discrimination differently across the $M_S$ range.

For larger values of the $y_{uu}$ Yukawa coupling, the signal yield greatly exceeds the background, leading to significantly smaller $p-$values compared to the more conservative case of $y_{uu}B(\chi \to Wb)=0.1$. The shorthand $y_{uu}=0.2(0.4)$ is used to denote $y_{uu}B(\chi\to Wb)=0.1(0.2)$.

This effect is illustrated in \Cref{fig:pval_yuu}, where larger $y_{uu}$ values lead to enhanced significances across most of the $M_S$ range, reaching $\approx3\sigma$.
This trend is driven by the increase in the $S_{uu}\to\chi\chi$ production cross section with $y_{uu}B(\chi\to Wb)$, consistent with Table VII from Ref.~\cite{Duminica_2025}.

The combined effect of increasing $y_{uu}$ and reducing the PDF uncertainties by a factor of two results in a noticeable improvement in the overall sensitivity, as shown in \Cref{fig:pval_yuu}, with discovery-level significance exceeding $5\sigma$ at $M_S=7.25~\text{TeV}$. In this case, the $M_S = 7~\text{TeV}$ point is omitted from the figure, as the corresponding $p-$value could not be reliably estimated with toy-based method, with no signal-like fluctuations observed even in samples of up to $10^9$ pseudo-experiments. However, the asymptotic Asimov formula of Ref.~\cite{Cowan_2011}, including background uncertainties, provides a complementary estimate, yields $p_0=3.1\cdot 10^{-11}$ for $M_S=7$ TeV, corresponding to $\approx 6.5\sigma$.

In summary, the analysis demonstrates that the sensitivity of the search is strongly driven by the product $y_{uu} B(\chi\to Wb)$. While large values lead to overwhelming signal dominance, realistic discovery or exclusion scenarios are better constrained by intermediate coupling strengths, where both signal and background contributions remain statistically meaningful. In this regime, PDF uncertainties can have a non-negligible impact on the achievable sensitivity. This interplay defines the region of parameter space most relevant for future searches at the HL-LHC.

\subsection{\label{subsec:upperlim}$CL_{s}$ scans and $95\%$ upper limits}

In addition to the discovery potential analysis based on local $p$-values, we perform an independent statistical study to determine the exclusion limits of the model through the $CL_s$ method and the 95\% confidence level (C.L.) upper limits on the signal strength parameter. 

This approach follows the standard procedures established by the ATLAS and CMS Collaborations~\cite{CMS-NOTE-2011-005,Read_2002}, providing a complementary perspective on the search sensitivity. The $CL_s$ technique combines the background-only and signal-plus-background hypotheses to assess the data compatibility with different signal strength values, enabling a consistent determination of exclusion regions in the $(M_S,y_{uu})$ parameter space. Besides evaluating the performance of the model when varying different parameters, the $CL_s$ and the upper limit quantify the sensitivity of the model to observing the $S_{uu}$ signal in a certain mass region.

\begin{figure}[h]
\centering
\includegraphics[width=0.5\textwidth]{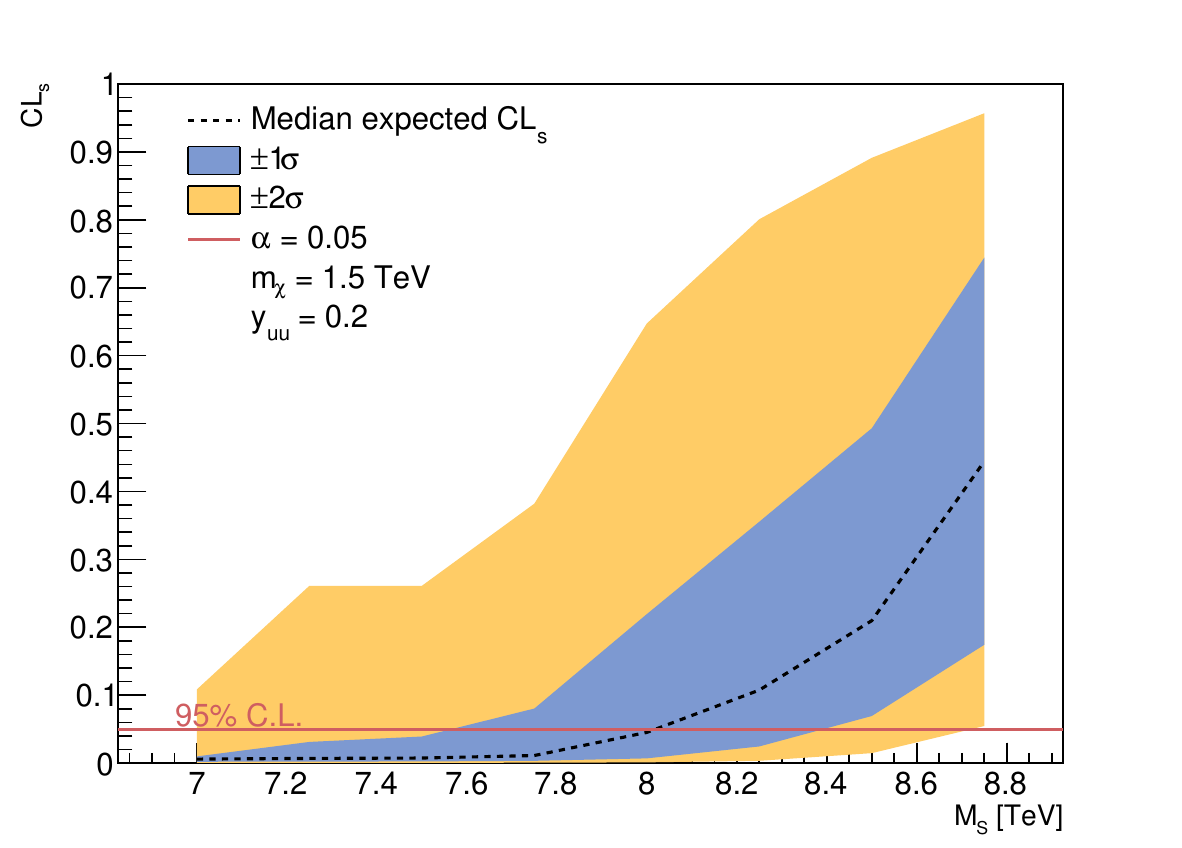} 
\caption{$CL_s$ scan in terms of $M_S$ for a VLQ mass of 1.5 TeV.}
\label{fig:chi15-cls}
\end{figure}

\begin{figure}[h]
\centering
\includegraphics[width=0.5\textwidth]{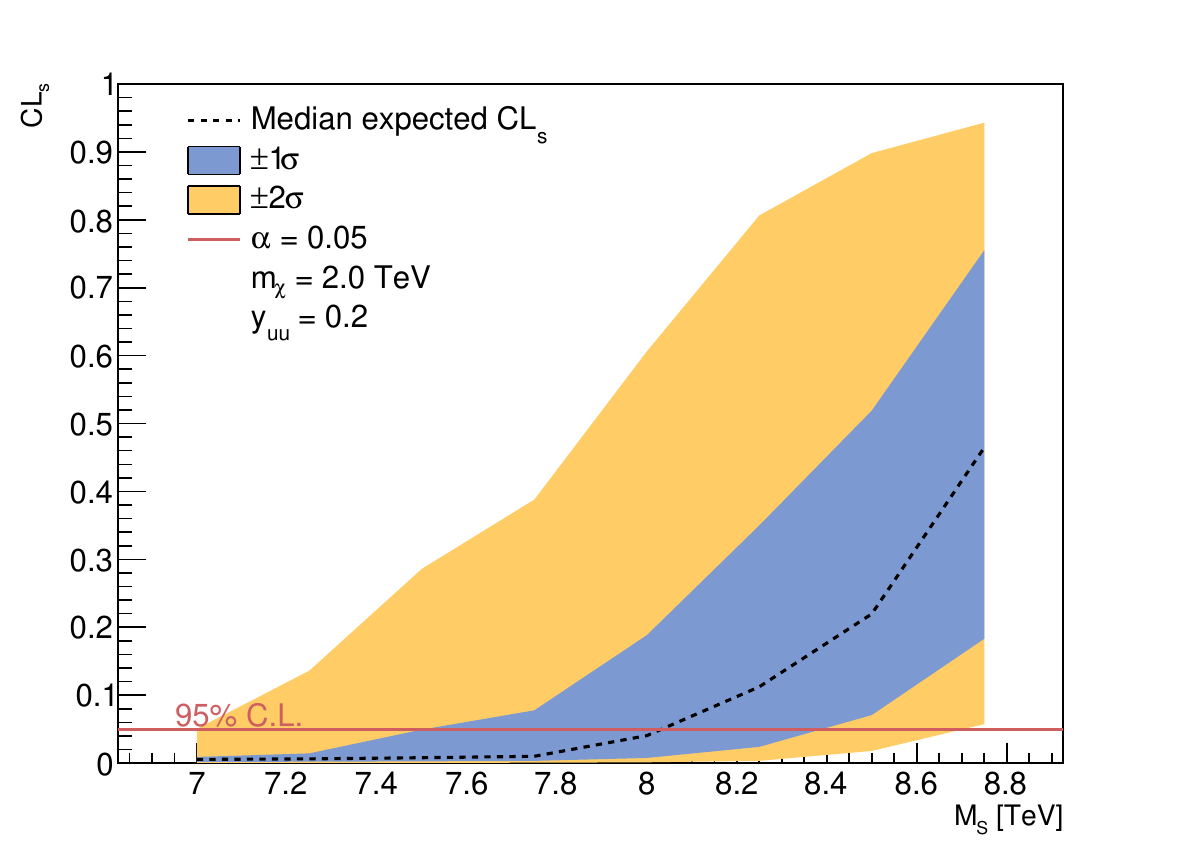} 
\caption{$CL_s$ scan in terms of $M_S$ for a VLQ mass of 2~TeV.}
\label{fig:chi2-cls}
\end{figure}

The results of the $CL_s$ scans for $m_\chi = 1.5~\text{TeV}$ and $m_\chi = 2~\text{TeV}$ are shown in \Cref{fig:chi15-cls,fig:chi2-cls}.


The intersection of the median expected $CL_s$ curve with the 95\% confidence level threshold defines the exclusion limits on the $S_{uu}$ mass. For both $m_\chi = 1.5~\text{TeV}$ and $m_\chi = 2~\text{TeV}$, the intersections occur at $M_S \approx 8~\text{TeV}$, showing that the analysis has comparable performance in both VLQ mass hypotheses. These values correspond to the expected 95\% confidence level exclusion limits, as determined from the $CL_s$ criterion described in \Cref{sec:stats}. The obtained limits are consistent with the recent CMS results reported in Ref.~\cite{CMS:2025hpa,CMS-PAS-EXO-25-004,Zisopoulos:2946439}.

\begin{figure}[h]
\centering
\includegraphics[width=0.5\textwidth]{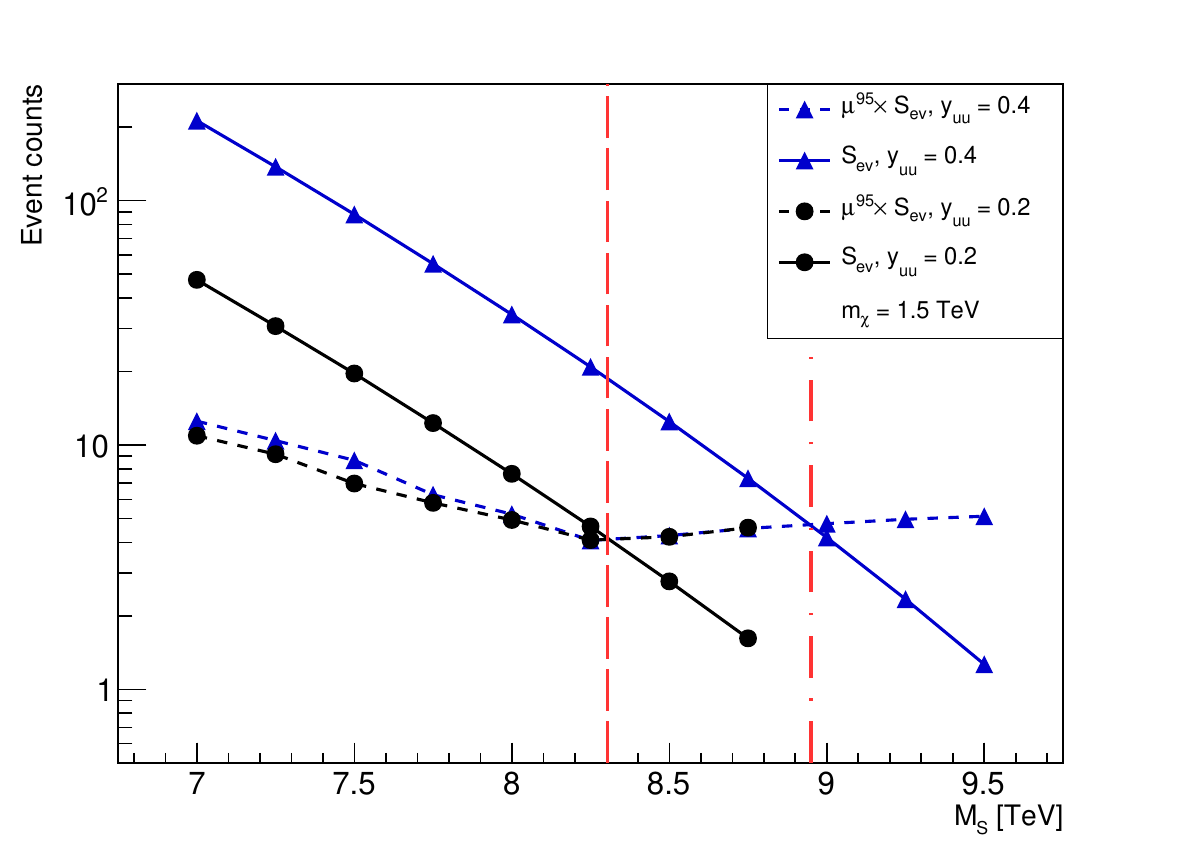} 
\caption{The theoretical signal yield $S_{\text{ev}}$ and the 95\% C.L. upper limit on $\mu\times S_{\text{ev}}$ for $M_S\in[7,~9.5]$ TeV, $m_\chi=1.5$ TeV, $D=0.90$ and $y_{uu}=0.2, \ 0.4$.}
\label{fig:chi15-ul}
\end{figure}

\begin{figure}[h]
\centering
\includegraphics[width=0.5\textwidth]{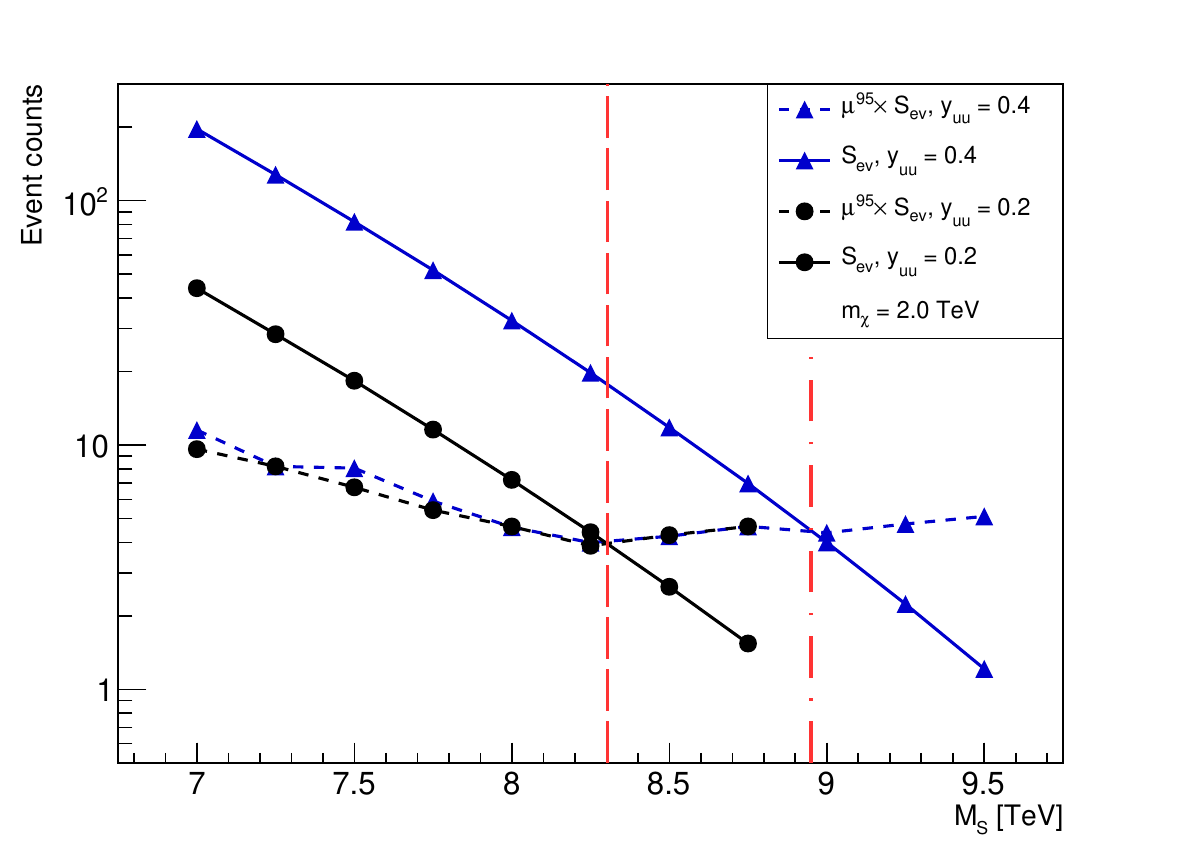}
\caption{The theoretical signal yield $S_{\text{ev}}$ and the 95\% C.L. upper limit on $\mu\times S_{\text{ev}}$ for $M_S\in[7,~9.5]$ TeV, $m_\chi=2$ TeV, $D=0.90$ and $y_{uu}=0.2, \ 0.4$.}
\label{fig:chi2-ul}
\end{figure}

\Cref{fig:chi15-ul,fig:chi2-ul} show the expected signal yields ($S_{\text{ev}}$) and the 95\% C.L. upper limits on $\mu^{95}\times S_{\text{ev}}$ as functions of $M_S\in[7,~9.5]~\text{TeV}$. The scans are performed for a discriminator value of $D = 0.90$ and two coupling scenarios, $y_{uu} = 0.2$ and $0.4$, and for $m_\chi = 1.5$ and $2~\text{TeV}$. The intersection points between the $S_{\text{ev}}$ and $\mu^{95} \times S_{\text{ev}}$ curves determine the expected exclusion limits on $M_S$. 


For $y_{uu} = 0.2$, the intersections occur at $M_S \approx 8.3~\text{TeV}$ for both $m_\chi = 1.5~\text{TeV}$ and $m_\chi = 2~\text{TeV}$. At $y_{uu} = 0.4$, the intersections can be observed at $M_s \approx 9~\text{TeV}$. The trend indicates that larger Yukawa couplings extend the accessibility $M_S$ range, while increasing the vectorlike quark mass poorly impacts the sensitivity, in alignment with the small differences in $S_{uu}\rightarrow\chi \chi$ production cross sections, for $m_\chi=1.5,~2~\text{TeV}$ (see \Cref{fig:cross-section}). The two statistical approaches - the toy-based $CL_s$ scan and the asymptotic upper limit computation - yield consistent results for the sensitivity of the $S_{uu}$ model. The differences are small and remain within the $1\sigma$ range, with both methods exhibiting the same overall trends as functions of $m_\chi$ and $y_{uu}$.

\begin{figure}[ht]
\centering
\includegraphics[width=0.5\textwidth]{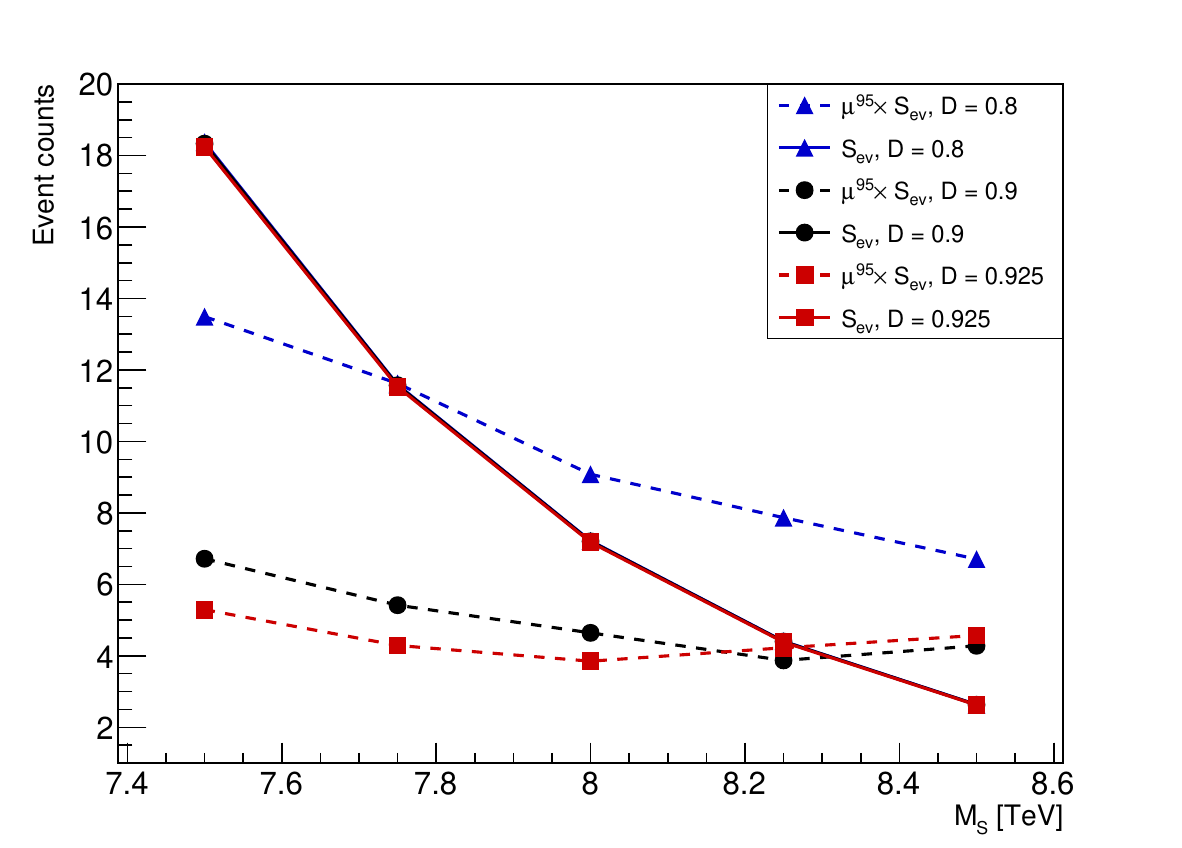} 
\caption{$S_{\text{ev}}$ and $\mu^{95}\times S_{\text{ev}}$ for $m_\chi=2$ TeV and $y_{uu}=0.2$ at $D=0.80,~ 0.90$ and 0.925. All three $S_{\text{ev}}$ curves are overlapping.}
\label{fig:chi20-ul-Disc}
\end{figure}

A complementary evaluation is carried out for the same values of the VLQ masses and Yukawa couplings at different ML discriminator thresholds, as shown in \Cref{fig:chi20-ul-Disc}.

The purpose of this comparison is to assess the variation of the exclusion limits under the concurrent increase in signal purity and decrease in event yields. As observed, within the range $D \in [0.80, 0.925]$, the intersection point between the $S_{\text{ev}}$ and $\mu^{95} \times S_{\text{ev}}$ curves shifts by $\approx0.5$~TeV during the transition from $D=0.80$ to $D=0.90$, while the change to $D=0.925$ presents a shift of only $0.1$~TeV. Considering the large uncertainties associated with the detector simulation, the parton distribution functions and jet reconstruction effects, this variation can be considered insignificant, especially in the case of the higher discriminator thresholds. At $D=0.80$ we gain a sizable fraction of background events, leading to a worse sensitivity of the model. These effects might be reduced with future improvements in the measurements.

In summary, the exclusion analysis based on both the toy–based MC and asymptotic $CL_s$ methods provides consistent and robust limits on the $S_{uu}$ mass across different parameter configurations. The expected 95\%~C.L. exclusion limits are found to vary between approximately 8 and 9 TeV, depending on the assumed vectorlike quark mass, Yukawa couplings, and ML discriminator threshold. Larger $y_{uu}$ values extend the $S_{uu}$ mass constraints, while higher $m_\chi$ values have a minimal influence on $M_S$ sensitivity, in agreement with the corresponding cross section behavior. The overall consistency of the two statistical approaches and the limited impact of the ML discriminator confirm the stability and reliability of the exclusion limits obtained for the $S_{uu}$ model at the HL-LHC experiments.

\section{Conclusions}\label{sec:conclusions} 

Extending our previous study~\cite{Duminica_2025}, this work presents a likelihood-based statistical analysis of the discovery and exclusion prospects for an ultraheavy diquark scalar $S_{uu}$ decaying into a pair of vectorlike quarks at the High-Luminosity LHC~\cite{ZurbanoFernandez:2020cco}. The analysis targets the fully hadronic six-jet final state from the process $pp \to S_{uu} \to \chi\chi \to (W^{+}b)(W^{+}b) \to (jjb)(jjb)$.

The results extend those of our earlier phenomenological study by providing a quantitatively controlled statistical interpretation. Sensitivity to the $S_{uu}$ resonance is observed in the mass range $8-9~\text{TeV}$, depending on the vectorlike quark mass, Yukawa couplings, and ML discriminator threshold. Larger values of $y_{uu}$ enhance the production cross section and extend the accessible mass range, while heavier vectorlike quarks leave the sensitivity nearly unchanged.

These results demonstrate that diquark-induced production of vectorlike quarks provides a viable and experimentally accessible probe of multi-TeV new physics at the HL-LHC within a fully hadronic final state.

The good agreement between the toy-based $CL_s$ and asymptotic approaches demonstrates the robustness of the statistical framework, while the limited dependence on the ML discriminator confirms the stability of the extracted exclusion limits.

Although the final states considered here differ from those analyzed by CMS, the results are in overall agreement with the exclusion ranges and trends reported in recent CMS publications~\cite{CMS:2025hpa,CMS-PAS-EXO-25-004,Zisopoulos:2946439}.

The main advancement of this work is the implementation of a consistent likelihood-based framework, including the treatment of theoretical and systematic uncertainties, enabling a robust and quantitative determination of discovery significance and exclusion limits.

Future studies will extend this framework to include complementary channels, such as $S_{uu} \to u\chi$ \cite{Filip:2601.11181} and $S_{uu} \to \chi\chi$ with $\chi\to tZ, ~th^0$, enabling a combined interpretation once correlated systematics and channel-dependent selections are consistently incorporated.

\section*{Acknowledgments}
We thank Bodgan Dobrescu, Stefan Ghinescu and Julien Maurer for valuable comments and suggestions.

The work of D.-C.C., C.A., A.-M.D, I.-M.D., I.D. and M.-S.F. was supported by IFIN-HH under Contract No.~PN-23210104 with the Romanian Ministry of Education and Research. The work of G.M. was supported by the project "Romanian Hub for Artificial Intelligence - HRIA", MySMIS no.~334906.

\providecommand{\noopsort}[1]{}\providecommand{\singleletter}[1]{#1}%

\end{document}